\documentclass[12pt]{article}

\newcommand{\bp}{{\bf P}}

\newcommand{\beq}{\begin{equation}}
\newcommand{\eeq}{\end{equation}}

\title{EPR-Bohm experiment, interference of probabilities, and imprecision of time}
\author{Andrei Khrennikov\footnote{Supported in part by the EU Human
Potential Programme, contact HPRN--CT--2002--00279 (Network on
Quantum Probability and Applications) and Profile Math. Modelling
in Physics and Cogn. Sc. of V\"axj\"o University.} \\International Center for Mathematical
Modeling \\ in Physics and Cognitive Sciences,\\
University of V\"axj\"o, S-35195, Sweden\\
Email:Andrei.Khrennikov@msi.vxu.se}
\date{}
\begin{document}
\maketitle

\abstract{  We demonstrate that the EPR-Bohm probabilities can be easily obtained in the
classical (but contextual) probabilistic framework by using the formula of
interference of probabilities. From this point of view the EPR-Bell experiment
is just an experiment on interference of probabilities. We analyze the time
structure of contextuality in the EPR-Bohm experiment. The conclusion is that
quantum mechanics does not contradict to a local realistic model in which
probabilities are calculated as averages over conditionings/measurements for
pairs of instances of time $t_1< t_2.$ If we restrict our consideration only
to simultaneous measurements at the fixed
instance of time $t$ we would get contradiction with Bell's theorem. One of implications of this fact
might be the impossibility to define instances of {\it time with absolute precision}
on the level of the microscopic
realistic model.}

\medskip

{\bf 1. Introduction.}In a series of papers, see, e.g., [1] it was demonstrated that {\it interference
of probabilities}(which is always considered as one of the main distinguishing features of quantum mechanics, e.g. [2]) can be easily
derived in contextual statistical framework (without to appeal to the Hilbert space formalism).
In this note it is demonstrated that in this framework we also can easily get the EPR-Bohm
probabilities. Moreover, the $\cos$-form of the EPR-Bohm probabilities is a consequence of interference
of probabilities corresponding to different contexts.

{\bf 2. Contextual statistical model.}  Basic structures of the model are
physical {\it contexts} -- complexes of physical conditions.\footnote{In principle, the notion of
context can be considered as a generalization of a widely used notion of
{\it preparation procedure}. } We denote the set of all
contexts by the symbol ${\cal C}.$

Suppose that there is fixed a set of observables
${\cal O}$  such that any observable $a\in {\cal O}$ can be measured
under a complex of physical conditions $C$ for any $C \in {\cal C}.$
We shall denote observables by Latin letters, $a,b,c,...,$ and their values by Greek letters,
$\alpha, \beta,\gamma, ...$

We do not assume that all these observables can be measured simultaneously;
so they need not be compatible. The sets of observables $ {\cal O}$ and contexts ${\cal C}$ are coupled
through

\medskip

{\bf Axiom 1:} {\it For  any observable
$a \in {\cal O},$ there are well defined contexts $C_\alpha$
corresponding to $\alpha$-filtrations: if we perform a measurement of $a$ under
the complex of physical conditions $C_\alpha,$ then we obtain the value $a=\alpha$ with
probability 1. It is supposed that the set of contexts ${\cal C}$ contains filtration-contexts $C_\alpha$
for all observables $a\in {\cal O}.$}

\medskip

{\bf Axiom 2:} {\it There are defined contextual probabilities ${\bf P}(a=\alpha/C)$ for any
context $C \in {\cal C}$ and any observable $a \in {\it O}.$}

{\bf 3. Interference of probabilities.}
Consider two dichotomous random
variables $a=\pm 1, b=\pm 1.$ Let $C$ be a context. There are well defined probabilities:
\[p_C^a(\alpha)={\bf P}(a=\alpha/C), p_C^b(\beta)={\bf P}(b=\beta/C),
p^{b/a}(\beta/\alpha)={\bf P}(b=\beta/a=\alpha),\]
We also introduce the matrix of transition probabilities:
${\bf P}^{b/a}=( p^{b/a}(\beta/\alpha)).$
The classical formula of total probability has the form:
$$
p_C^b(\beta)= p_C^a(+) p^{b/a}(\beta/+) + p_C^a(-) p^{b/a}(\beta/-) .
$$
However, for contextual probabilities this formula can be violated, see [1]. We introduce the {\it coefficient
of statistical incompatibility} of observables $a$ and $b$ in the context $C$ (see [1]):
$$
\lambda(\beta/a, C)=\frac{p_C^b(\beta) - [p_C^a(+) p^{b/a}(\beta/+) + p_C^a(-) p^{b/a}(\beta/-)]}
{2  \sqrt{p_C^a(+)p^{b/a}(\beta/+)p_C^a(-) p^{b/a}(\beta/-)}}
$$
and by using this coefficient we can rewrite the probability $p_C^b(\beta)$ in the interference-like
form:
$$
p_C^b(\beta)= p_C^a(+) p^{b/a}(\beta/+) + p_C^a(-) p^{b/a}(\beta/-) +
$$
$$
2 \lambda(\beta/a, C) \sqrt{p^{b/a}(\beta/+)p_C^a(-) p^{b/a}(\beta/-)}.
$$
Now let us restrict our considerations to the case of relatively small coefficients of incompatibility:
$$
\vert \lambda(\beta/a, C)\vert \leq 1.
$$
In this case we can introduce new statistical parameters $\theta(\beta/a, C)\in
[0,2 \pi]$ and represent the coefficients of incompatibility in the
trigonometric form:
$$
\lambda(\beta/a, C)=\cos \theta (\beta/a, C).
$$
Parameters $\theta(\beta/a, C)$ are said to be {\ statistical phases.}
For such contexts we get the standard formula of interference of probabilities:
$$
p_C^b(\beta)= p_C^a(+) p^{b/a}(\beta/+) + p_C^a(-) p^{b/a}(\beta/-) +
$$
\begin{equation}
\label{INT}
2 \cos(\beta/a, C) \sqrt{p^{b/a}(\beta/+)p_C^a(-) p^{b/a}(\beta/-)}
\end{equation}
which is usually derived by using the Hilbert space formalism.\footnote{If the coefficients of incompatibility
are larger than 1, we obtain so called hyperbolic interference [1] which is described by hyperbolic quantum mechanics
[3]. We do not consider this possibility in the present paper, but we notice that both ordinary (``trigonometric'')
quantum mechanics and hyperbolic quantum mechanics can be obtained as deformations of the same
classical mechanics [3].}

{\bf Important remark}  Quantum observables always produce {\it double stochastic matrices}
of transition probabilities, i.e., $p^{b/a}(+/\alpha) + p^{b/a}(-/\alpha)=1$ and
$p^{b/a}(\beta/+) + p^{b/a}(\beta/-)=1.$

{\bf 4. EPR-Bohm probabilities as interference probabilities.}
Let us now consider three dichotomous  observables $a, b, c=\pm 1.$ There are well defined
selection-contexts $C_\pm,$ selections with respect to values $\gamma=\pm 1.$
We choose $C=C_+$ (i.e., the context corresponding to the $c=+1$ selection) and use results of section 3
for this context. There are well defined probabilities
$$
p_C^a(\alpha)\equiv p(a=\alpha/C), \alpha=\pm 1, \; p_C^b (B)\equiv p(b=\beta/C), \beta=\pm 1.
$$
By taking into account that $C=C_+$ we get:
$$
p_C^a(\alpha)=p^{a/c}(\alpha/+), \; p_C^b(\beta)\equiv p^{b/c}(\beta/+).
$$
We now can use the formula of total probability with interference term, (\ref{INT}), to the $b$-observable.
 For $\beta=+1$ we have:
\[p^{b/c}(+/+)=p^{a/c}(+/+) p^{b/a}(+/+) + p^{a/c}(-/+) p^{b/a}(+/-)+\]
\[ 2\cos \theta_+\sqrt{p^{a/c}(+/+) p^{b/a}(+/+) p^{a/c}(-/+) p^{b/a}(+/-)},\]
where $\theta_+ = \theta (+/a, C_+).$ We can always represent probabilities in the trigonometric form:
$$
p^{a/c}(+/+)=\cos^2 \xi, \; \; p^{a/c}(-/+)=1-\cos^2 \xi=\sin^2 \xi,
$$
where $\xi=\xi^{a/c}(+/+).$
Suppose now that matrix$\bp^{b/a}$ is double stochastic.
Then we can represent probabilities:
 $$
 p^{b/a}(+/+)=\sin^2 \eta, \;   \;
 p^{b/a}(+/-)=1-\sin^2 \eta=\cos^2 \eta,
 $$
 where $\eta=\eta^{a/b}(+/+).$ To simplify considerations,
 we consider the case when both phases  $\xi, \eta, \in (0, \pi/2).$
 Thus we have:
 $$
 p^{b/c}(+/+)=\cos^2 \xi \sin^2 \eta + \sin^2 \xi \cos^2 \eta +
 2 \cos \theta_+ \cos \xi \sin \eta \sin \xi \cos \eta.
 $$
 We now perform similar considerations for $\beta=-1:$
 \[p^{b/c}(-/+)=p^{a/c}(+/+)p^{b/a}(-/+)+p^{a/c}(-/+)p^{b/a}(-/-)+\]
 \[2\cos \theta_-\sqrt{p^{a/c}(+/+) p^{b/a}(-/+) p^{a/c}(-/+) p^{b/a}(-/-)},\]
 where $\theta_-=\theta(-/a, C_+).$
 We have  that
 $$
 p^{b/a}(-/+)=1-p^{b/a}(+/+)=\cos^2 \eta, \; \;
 p^{b/a}(-/-)=1-p^{b/a}(+/-)=\sin^2 \eta.
 $$
 Thus
 $$
 p^{b/c}(-/+)=\cos^2 \xi \cos^2 \eta + \sin^2 \xi \sin^2 \eta + 2 \cos \theta_- \cos \xi \cos \eta \sin \xi \sin \eta
 $$

 Consider now the case of trigonometric interference the {\it maximal magnitude}:
  \beq
 \label{MAX}
 |\lambda_\pm|=|\cos \theta_\pm|=1
 \eeq

 {\bf Lemma 1.}{\it{ Let $\bp^{b/a}$ be double stochastic and let the condition (\ref{MAX}) hold. Then}}

 \beq
 \label{OP}
 \cos \theta_+=-\cos \theta_-
 \eeq

 {\bf Proof.}
 We have

 \[1=p^{b/c}(-/+)+p^{b/c}(+/+)=\]
 \[\cos^2 \xi (\cos^2 \eta + \sin^2 \eta) + \sin^2 \xi (\cos^2 \eta + \sin^2 \eta)+\]
 \[2\cos \theta_+ \cos \xi \sin \eta \sin \xi \cos \eta + 2 \cos \theta_- \cos \xi \cos \eta \sin \xi \sin \eta\]

 This implies (\ref{OP}) (since $\xi, \eta \in (0, \pi/2)).$

\medskip

 Let us say that
 \beq
 \label{OP1}
 \cos \theta_+=-1, \cos \theta_-=+1
 \eeq
 (we can make the opposite choice but this would induce a phase shift). Then we get
 $$
 p^{b/c}(+/+)=(\cos \xi \sin \eta - \sin \xi \cos \eta)^2=
 \sin^2 (\xi - \eta),
 $$
 $$
 p^{b/c}(-/+)=(\cos \xi \cos \eta + \sin \xi \sin \eta)^2=\cos^2 (\xi-\eta)
 $$
 We now consider representations of probabilities $p^{b/c}(\pm/-)$ for the context $C=C_-$
 (selection for the value $c=-1).$
 \[ p^{b/c}(+/-)=p^{a/c}(+/-)p^{b/a}(+/+)+p^{a/c}(-/-)p^{b/a}(+/-)+\]
 \[2\cos \tilde \theta_+ \sqrt{p^{a/c}(+/-)p^{b/a}(+/+)p^{a/c}(-/-)p^{b/a}(+/-)},\]
 where $\tilde \theta_+\equiv \theta(+/a, C_-).$
  Suppose that the matrix $\bp^{a/c}$ is also double stochastic. Then  we get
 $$
 p^{a/c}(+/-)=1-\cos^2 \xi=\sin^2 \xi, \;\; p^{a/c}(-/-)=\cos^2 \xi.
 $$
 So we have:
 $$
 p^{b/c}(+/-)=\sin^2 \xi \sin^2 \eta + \cos^2 \xi \cos^2 \eta +
 2\cos \tilde \theta_+ \sin \xi \sin \eta \cos \xi \cos \eta.
 $$
 In the same way we get:
 $$
 p^{b/c}(-/-)=\cos^2 \xi \sin^2 \eta + \sin^2 \xi \cos^2 \eta +
 2 \cos \tilde \theta_- \cos \xi \sin \eta \sin \xi \cos \eta.
 $$

 {\bf Lemma 2.}{\it{ Let matrices $\bp^{b/a}, \bp^{a/c}, \bp^{b/c}$ be double stochastic
 and have strictly positive elements. Then}}
 \beq
 \label{EQ}
 \cos \theta_+=-\cos\tilde \theta_+, \cos \theta_-=-\cos \tilde \theta_-
 \eeq

{\bf Proof.} By using double stochasticity of $\bp^{b/c}$ we get
$$
1=p^{b/c}(+/+)+p^{b/c}(+/-)=\sin^2 \eta (\cos^2 \xi + \sin^2 \xi) +
$$
$$
\cos^2 \eta (\cos^2 \xi + \sin^2 \xi) + 2(\cos \theta_+ + \cos \tilde \theta_+) \cos \xi \cos \eta \sin\xi \sin\eta.
$$
By using that $\xi, \eta \in (0, \frac{\pi}{2})$
(this we can always assume since \\ $p^{b/a}(\beta/\alpha), p^{a/c}(\alpha/\gamma)>0),$
we get $\cos \theta_+=-\cos \tilde \theta_-.$

Results of all these considerations and computations can be represented as

 {\bf{Theorem.}}{\it{ Let all matrices $\bp^{b/a}, \bp^{a/c}, \bp^{b/c}$ be double
 stochastic and let interference parameter $\lambda_\pm=\cos \theta_\pm$
 have the maximal absolute magnitude, $|\lambda_\pm|=1.$ Then
 probabilities can be represented in the EPR-Bohm form:}}
 $$
 p^{b/c}(+/+)=p^{b/c}(-/-)=\sin^2 (\xi-\eta)
 $$
 \beq
  \label{EB}
 p^{b/c}(+/-)=p^{b/c}(-/+)=\cos^2 (\xi-\eta)
 \eeq

\medskip

 {\bf Conclusion.} {\it In the contextual approach the EPR-Bohm probabilities can be interpreted as interference
 probabilities.}

{\bf 5. Physical consequences.} What are main physical consequences of
the contextual probabilistic derivation of the EPR-Bohm probabilities?

The evident consequence is that
{\it local realism is compatible with quantum mechanics}\footnote{We remark that we do not consider
Bell's inequality.} if contextuality of probabilities is taken into account.

However, to make physicists interested in our contextual approach to the EPR-Bohm experiment,
we should be able to find the physical mechanism of contextuality in this experiment. We recall
that contextuality
is dependence on complexes of physical conditions. To obtain the EPR-Bohm probabilities in our approach,
we should first make selection with respect to one fixed context, e.g.,  the context
$C=C_+$ corresponding to the selection of subensemble of pairs such that the measurement of the
$c$-projection of spin (or polarization) on the first particle in a pair gives the value $c=+1.$
We emphasize that, despite the fact the measurement is performed only on the first particle in  a pair
of correlated particles, the selection $C_+$ is selection of pairs. Such a selection creates a new ensemble
and on this new ensemble we perform measurements of $a$ or $b$-projections for the second particle in a pair.
In this way we obtain the same probabilities which are predicted by quantum mechanics (by using the formalism
of complex Hilbert space).  The crucial point is that the order structure -- contextual
selection and then measurement -- should always be taken into account. A selection performed at some instant
of time $t_1$ gives an ensemble which is used for measurement at some instant of time $t_2 > t_1.$  Thus, to have
a local realistic picture, we should assume that the contextual selection and measurement are performed in different instances of time.
We remark that it is not important on which of particles (on the first or on the second)
there is performed the selection measurement and on which the final measurement. Probabilities
are symmetric with respect to these procedures. This is a consequence of double stochasticity.

Thus in our model the EPR-Bell probabilities
are obtained as the result of the average of contextual $t_2/t_1$-probabilities over all pairs
of instances of time $t_1 < t_2.$

The  model cannot say anything about simultaneous measurements,
$t_1=t_2.$  However, the contribution of simultaneous measurements,
$t_1=t_2,$ is negligibly small, since the measure of the diagonal in any two dimensional
time-rectangle equals to zero.

If we restrict our consideration only to simultaneous measurements at the fixed
instance of time $t$ we would get contradiction with Bell's theorem. One of implications of this fact
might be the impossibility to define instances of {\it time with absolute precision}
on the level of the  microscopic
realistic model . Such a conjecture is quite natural
if we take into account time-energy uncertainty relation. This relation (derived in quantum formalism),
of course, should be also valid for the prequantum local realistic model.

{\bf References}

 1. A. Yu. Khrennikov, {\it J. Phys.A: Math. Gen.,} {\bf 34}, 9965-9981 (2001);{\it J. Math. Phys.},
{\bf 44}, 2471- 2478 (2003); {\it Phys. Lett. A}, {\bf 316}, 279-296 (2003).

2. P. A. M.  Dirac, {\it The Principles of Quantum Mechanics}
(Claredon Press, Oxford, 1995).

R. Feynman and A. Hibbs, {\it Quantum Mechanics and Path Integrals}
(McGraw-Hill, New-York, 1965).

3. A. Yu. Khrennikov,  {\it Annalen  der Physik,} {\bf 12}, 575-585 (2003); quant-ph/0401035.

\end{document}